\documentclass[a4paper,twocolumn,11pt]{quantumarticle}
\pdfoutput=1
\usepackage[utf8]{inputenc}
\usepackage[english]{babel}
\usepackage[T1]{fontenc}
\usepackage{amsmath}
\usepackage{hyperref}
\usepackage{float}
\usepackage{tikz}
\usepackage{lipsum}
\usepackage{enumitem}

\usepackage{listings}
\usepackage{lstautogobble}
\usepackage{amssymb}
\usepackage{xcolor}
\usepackage{bracketkey}

\newcommand{\bc}{\begin{center}}
\newcommand{\ec}{\end{center}}
\newcommand{\be}{\begin{equation}}
\newcommand{\ee}{\end{equation}}
\newcommand{\ba}{\begin{array}}
\newcommand{\ea}{\end{array}}
\newcommand{\beq}{\begin{eqnarray}}
\newcommand{\eeq}{\end{eqnarray}}
\newcommand{\ket}[1]{\left| {#1}\right\rangle}
\newcommand{\bra}[1]{\left\langle {#1} \right|}

\newcommand{\abs}[1]{\left| {#1}\right|}

\begin{document}


\title{Qurrium: A Python package for randomized measurement-based estimation of quantum state properties}

\author{Huai-Chun Chang}
\orcid{0009-0005-7526-4419}
\affil{Research Center for Critical Issues, Academia Sinica, Guiren, Tainan 711010, Taiwan}
\affil{Graduate Institute of Applied Physics, National Chengchi University, Taipei 11605, Taiwan}
\author{Teik-Hui Lee}
\orcid{0000-0002-8220-1396}
\affil{Research Center for Critical Issues, Academia Sinica, Guiren, Tainan 711010, Taiwan}
\author{Arthur Strauss}
\affil{Centre for Quantum Technologies, National University of Singapore, Singapore}
\affil{Quantum Machines, Singapore}

\author{Yu-Cheng Lin$^*$}
\orcid{0000-0001-6112-3723}
\email{yc.lin@nccu.edu.tw}
\affil{Graduate Institute of Applied Physics, National Chengchi University, Taipei 11605, Taiwan}
\author{Hsiu-Chuan Hsu$^*$}
\orcid{0000-0002-8295-9092}
\email{hcjhsu@nccu.edu.tw}
\affil{Graduate Institute of Applied Physics, National Chengchi University, Taipei 11605, Taiwan}
\affil{Department of Computer Science, National Chengchi University, Taipei 11605, Taiwan}



\begin{abstract}
Estimating quantum state properties is essential across a wide range of
applications, from studying quantum many-body physics to benchmarking quantum
hardware. We present Qurrium, a Python package built on Qiskit that implements
randomized measurement protocols for estimating purity, second-order R\'{e}nyi
entropy, expectation values of Pauli observables, and state overlap. In this
paper, we focus on the classical shadow protocol and demonstrate two workflows
in Qurrium. One is the end-to-end workflow that integrates quantum circuit
preparation, simulation, measurement, and analysis. The other is the standalone
estimation workflow that accepts pre-collected measurement data from any
hardware platform in Qurrium's data format. We demonstrate both workflows
through examples of a cluster state and an Ising time-evolved state.
Furthermore, we report results obtained from a superconducting quantum
processor developed by Academia Sinica and analyzed using the standalone
estimation workflow. Qurrium is built on Qiskit, the dominant framework in
quantum computing software, making it directly accessible to the large community of researchers already working with Qiskit. The source code is openly available at \url{https://github.com/qurrium/qurrium} and the example code is provided at \url{https://github.com/qurrium/classical-shadow-examples}.
\end{abstract}

\section{Introduction}
\label{sec:intro}
Characterizing quantum states is essential for a wide range of applications in
quantum computing, including Hamiltonian simulation, hardware calibration,
benchmarking, and algorithm design. 
Randomized measurement protocols~\cite{Elben2019,Elben2020,Huang2020,Elben2023} provide a powerful and
resource-efficient framework for estimating quantum state properties, and have
been successfully demonstrated across a variety of hardware platforms, including
trapped ions~\cite{Brydges2019,Wilkens2026}, superconducting qubits~\cite{Vovrosh2021,
Yang2026}, photonic systems \cite{Zhang2021,Li2026}, and Rydberg
atoms~\cite{Notarnicola2023,Matsoukas2026}. As quantum hardware matures
\cite{Kim2023,Acharya2025,Chiu2025} and finds broader applications across disciplines such as
materials science, quantum chemistry~\cite{Daley2022} and
finance~\cite{Herman2023}, there is an increasing need for software that enables reliable estimation of quantum
state properties with a low barrier to entry. 
For example, a recently developed software package {\sf RandomMeas.jl}~\cite{Elben2025}, written in
Julia, provides a comprehensive implementation of randomized measurement
protocols.
The package incorporates {\sf ITensors.jl},
enabling the simulation of large-scale quantum systems beyond the reach of
state-vector methods. 

Since Qiskit and Python remain  dominant in the quantum computing software ecosystem~\cite{Upadhyay2025}, 
we develop a randomized measurement toolbox, {\sf Qurrium}, that provides a Python-native
implementation built directly on Qiskit. This design offers an accessible option for
the large community of researchers already working within the Qiskit ecosystem. 
The supported protocols in {\sf Qurrium} 
include the Haar-measure based randomized measurement (HM-RM)
protocol~\cite{Brydges2019, Elben2019,Elben2020,Mele2024} and the classical
shadow protocol~\cite{Huang2020} based on random Pauli measurements. The
package provides an end-to-end workflow encompassing quantum circuit execution and data analysis,
as well as a standalone estimation workflow for users with pre-collected
measurement data.

An earlier version of {\sf Qurrium}
focused on the HM-RM protocol and was applied to estimate the second-order
R\'{e}nyi entropy of quantum states on IBM Q hardware~\cite{Chang2025}. In this
work, we focus on the classical shadow protocol and demonstrate the capabilities of {\sf Qurrium}
through two use cases: a time-independent example based on cluster state and a
dynamical example involving Ising time evolution. We demonstrate the convenience of the
end-to-end workflow, including quantum circuit execution on
\texttt{AerSimulator} provided by Qiskit. We further highlight the
compatibility of the standalone estimation workflow with any hardware platform:
as long as measurement outcomes and the corresponding random unitaries are recorded in
the {\sf Qurrium} data format, the data can be directly processed by the package. 
As a hardware demonstration, we report results obtained using
the prototype superconducting qubit processor developed by Academia Sinica in Taiwan. 

The rest of the paper is organized as follows. Section~\ref{sec:overview} reviews
the classical shadow protocol. Section~\ref{sec:software} introduces the
workflows for the classical shadow protocol in {\sf Qurrium}, including the
end-to-end and standalone estimation workflows. Section~\ref{sec:ex} presents two
examples, a cluster state and dynamical states generated by Ising evolution, to
demonstrate the usage of these workflows.
Section~\ref{sec:concl} provides the conclusion and outlook.

\section{Overview of the classical shadow
protocol}\label{sec:overview} The classical shadow protocol~\cite{Huang2020}
provides a resource-efficient framework for estimating properties of a quantum
state from randomized measurements. One of the key advantages is that the
measurement settings need not be tailored to the observable of interest.
Instead, one performs randomized measurements first and determines the
quantities to be estimated in post-processing. This  paradigm is summarized as
\textit{measure first, ask questions later}~\cite{Elben2023}. 

After preparing an $n$-qubit quantum state $\rho$ using a quantum circuit, the
state is measured independently under $N_U$ random Pauli settings.  For each
setting, a random tensor-product unitary $U = \bigotimes_{i=1}^n U_i$ is
applied to the state, followed by measurement in the computational basis.  Each
single-qubit Pauli unitary $U_i$ satisfies $U_i^{\dagger} Z  U_i= P_{U_i} \in
\{X, Y, Z \}$.  In {\sf Qurrium} the ensemble $U_i \in \{H, HS^\dagger, I \}$
is the default set generating the Pauli bases. Each measurement setting is
repeated $N_K$ times, referred to as {\sf shots} in {\sf Qurrium} and in this paper.
For each measurement setting, the single-qubit unitaries $U_i$ applied to each qubit, 
together with the measurement outcomes, represented by bitstrings ${b}\in \{0, 1 \}^n$ and 
their corresponding counts, are recorded.

For a fixed setting $m$ with unitary $U^{(m)} = \bigotimes_{i=1}^n U_i^{(m)}$ and a
measured bitstring $b_k = (b_{k,1},\dots,b_{k,n})$ obtained on shot $k$, the
single-qubit snapshot (measurement projector) is defined as 
\be
\Pi_{k,i}^{(m)} = U_i^{\dagger(m)} \ket{b_{k,i}}\bra{b_{k,i}} U^{(m)}_i\,.
\ee
Applying the single-qubit inverse of the measurement channel $\mathcal{M}_1^{-1}$ on each qubit,
the classical shadow of a single shot $k$ in setting $m$ is given by
\be
\rho^{(m)}_k = \bigotimes_{i=1}^n \sigma_{k,i}^{(m)} = \bigotimes_{i=1}^n \left( 3 \Pi_{k,i}^{(m)}  -I  \right)\,.
\label{eq:rho_m}
\ee
Averaging over the $N_K$ shots collected for measurement setting $m$ yields
\be
\overline{\rho^{(m)}} = \frac{1}{N_K} \sum_{k=1}^{N_K} \bigotimes_{i=1}^n \left( 3 \Pi_{k,i}^{(m)}  -I  \right)\,,
\label{eq:shot_av}
\ee
here we use an overline $\overline{(\,\cdot\,)}$ to denote a shot average.
The set $\{\overline{\rho^{(m)}}\}_{m=1,\cdots,N_U}$ obtained from all
$N_U$ measurement settings is referred to as a classical shadow of the state
$\rho$~\cite{Huang2020}. 
The ensemble average over all $N_U$ settings (denoted by $[\,\cdot\,]_\text{av}$) yields the classical-shadow estimator of the density
matrix,
\be
\hat{\rho}\equiv [\,\overline{\rho}\,]_\text{av} = \frac{1}{N_U} \sum_{m=1}^{N_U} 
\overline{\rho^{(m)}}. 
\label{eq:ensemble_av}
\ee
which satisfies $\mathbb{E}[\hat{\rho}] = \rho$~\cite{Huang2020}.

To compute the expectation value of an operator $\mathsf{O}$ with respect to a state $\rho$,
the density matrix $\rho$ can be replaced by its corresponding classical shadow, yielding
\begin{equation}
        \langle\mathsf{O}\rangle = \mathrm{Tr}(\mathsf{O}\rho) = \mathrm{Tr}(\mathsf{O}\,\mathbb{E}[\hat{\rho}]) 
       \label{eq:EO}
\end{equation}

Beyond expectation values of linear observables, the classical shadow protocol also enables estimation of nonlinear functionals of the density matrix~\cite{Elben2023}. A prominent example is the purity of a subsystem $A$, defined as
\begin{equation}
	p_2 = \mathrm{Tr}(\rho_A^2),
\end{equation}
where $\rho_A = \mathrm{Tr}_B(\rho)$ is the reduced density matrix obtained by
tracing out the complement $B$ of $A$. When $A$ is the entire system, $\rho_A = \rho$
corresponds to the full density matrix. An unbiased estimator of the purity is given by
~\cite{Elben2023}
\begin{equation}
	\hat{p}_2 = \frac{1}{N_U(N_U-1)}\sum_{m\neq m'}
	\mathrm{Tr}(\overline{\rho_A^{(m)}}\,\overline{\rho_A^{(m')}}),
\end{equation}
where the sum runs over all pairs of distinct random measurement unitaries. 

Another extension of the classcial shadow protocol is the estimation of the overlap (or fidelity) $\text{Tr}(\rho \rho')$ 
between two distinct quantum states $\rho$ and $\rho'$. When the density matrix of the target state $\rho'$
is known explicitly, denoted by $\sigma$, the overlap can be directly estimated as 
\begin{equation}
        \hat{f}_1(\rho, \sigma) =
        \mathrm{Tr}(\sigma\,\mathbb{E}[\hat{\rho}]).
        \label{eq:linoverlap}
\end{equation}
For two states prepared independently using quantum circuits,
an unbiased estimator of their overlap is instead given by
\be
\hat{f}_2(\rho, \rho') = \frac{1}{N_U} \sum_{m=1}^{N_U}
\text{Tr}\bigl[\overline{\rho^{(m)}}\; \overline{\Pi'^{(m)}}  \bigr]\,,
\label{eq:f_overlap}
\ee
where $\overline{\rho^{(m)}}$ is the shot-averaged, {\it inverted} shadow of $\rho$
in  measurement setting $m$, while
\be
\overline{\Pi'^{(m)}}=\frac{1}{N_K} \sum_{k=1}^{N_K} \bigotimes_{i=1}^n {\Pi'}_{k,i}^{(m)}
\ee
is the shot-average of the {\it uninverted} (raw) measurement projectors for $\rho'$ under the same measurement
setting $m$. A derivation provided in the Appendix \ref{app:overlap} makes it explicit that this asymmetry (inverting one state but not the other) is precisely what ensures that Eq.~\eqref{eq:f_overlap} constitutes an unbiased estimator of $\mathrm{Tr}(\rho \rho')$.

\section{Software Design and Workflow}\label{sec:software}

 \begin{figure*}[t]
 	\centering
 	\includegraphics[scale=1]{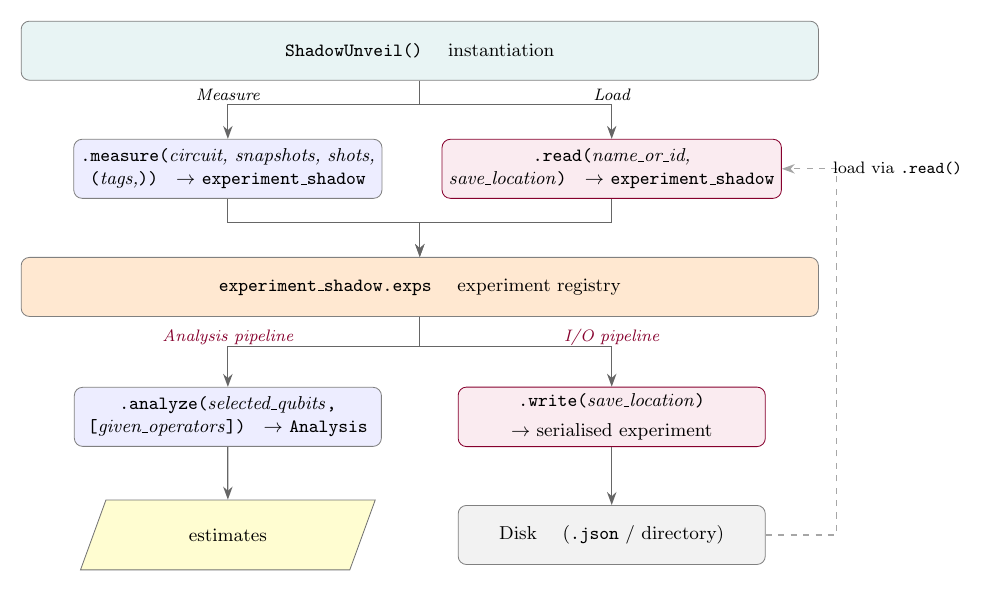}
 	\caption{The end-to-end workflow for the classical shadow protocol in {\sf Qurrium}.  
The figure was generated using the \texttt{TikZ} package in \LaTeX, written with the assistance of Claude Sonnet 4.6 (Anthropic).} 
 	\label{fig:e2e}
 \end{figure*}

In this paper, we highlight two workflows in {\sf Qurrium} for estimating quantum properties using classical shadow protocol, the end-to-end workflow and the standalone estimation workflow.
In the end-to-end workflow, as shown in Fig.~\ref{fig:e2e}, the user instantiates the \texttt{ShadowUnveil} class and invokes one of two entry-point methods. The \texttt{.measure()} method performs the randomized measurements by applying a product of single-qubit random unitaries to the quantum state and recording the measurement outcomes. Once all measurement outcomes are collected, \texttt{.analyze()} computes the quantum properties of interest returns the estimated values. For users who wish to save the full experiment and analysis within {\sf Qurrium}, the I/O pipeline provides convenient serialization of experiments to disk via \texttt{.write()} and reloading for re-analysis via \texttt{.read()}, the other entry-point methods. 

\begin{figure}[t]
	\centering
	\includegraphics[width=1\linewidth]{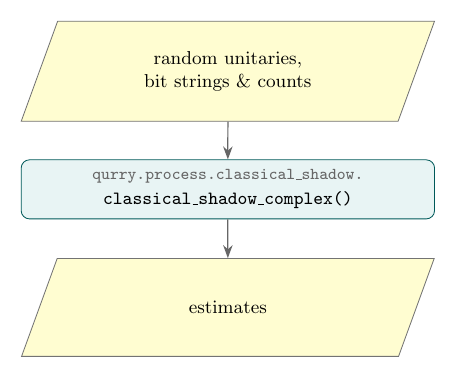}
	\caption{The stadnalone estimation workflow. The figure was generated using the \texttt{TikZ} package in \LaTeX, written with the assistance of Claude Sonnet 4.6 (Anthropic).} 
	\label{fig:standalone}
\end{figure}

In the standalone estimation workflow, as shown in Fig. \ref{fig:standalone}, the user supplies pre-collected measurement outcomes and random bases directly to 
\texttt{classical\_shadow\_complex}, available as
\begin{widetext}
\begin{lstlisting}[numbers=none]
	from qurry.process.classical_shadow import classical_shadow_complex
\end{lstlisting}
\end{widetext}
This workflow is designed for users who have already collected  measurement data, for example, from a real hardware experiment, and wish to perform the estimation without repeating the simulation stage. By default, \texttt{classical\_shadow\_complex} returns estimates of the density matrix and purity. Expectation values of user-supplied operators are additionally returned when provided.

\section{Illustrative examples}\label{sec:ex}

This section demonstrates how to use the workflows provided by
{\sf Qurrium} to prepare quantum states and characterize their properties.
We consider two examples: a cluster state and a dynamical state generated by
Ising time evolution. For each example, we demonstrate the calculation of the
purity, second-order R\'{e}nyi entropy, expectation values of Pauli operators, and
state overlap.

{\sf  Qurrium} adopts the little-endian bit and qubit ordering convention used by Qiskit,
where the first qubit (index 0) represents the least significant bit, 
placed in the rightmost position of a statevector ket or bitstring.
Throughout this section, qubit indices follow the standard zero-based
convention, ranging from $0$ to $n-1$. 
The companion code for all demonstrations is available at \url{https://github.com/qurrium/classical-shadow-paper-example}.


\subsection{Example: Cluster state}
\label{sec:cluster}

\begin{figure}[t]
        \centering
        \includegraphics[width=0.6\linewidth]{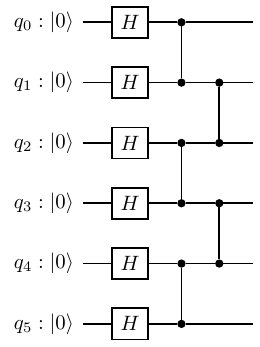}
        \caption{Quantum circuit for preparing a six-qubit one-dimensional cluster state with OBC. \label{fig:cluster}} 
\end{figure}

We first examine properties of the one-dimensional cluster state $\ket{C_n}$, defined as  
the unique ground state of the Hamiltonian~\cite{Choo2018}
\be
H_C = - \sum_{i=0}^{n-1} Z_{i-1} X_i Z_{i+1}\,,
\label{eq:Hc} 
\ee 
where we set  $Z_{-1}\equiv I$ and $Z_n\equiv I$ for systems with open boundary conditions (OBC),
and $Z_{-1}\equiv Z_{n-1}$ and $Z_n\equiv Z_0$ for periodic boundary conditions (PBC).
Since all terms in $H_C$ commute, $\ket{C_n}$ satisfies the stabilizer conditions for every site $i$:
\be
g_i \ket{C_n} = \ket{C_n},\quad \text{with } g_i= Z_{i-1} X_i Z_{i+1}\,.
\ee

The $n$-qubit cluster state can be written through the circuit construction
explicitly as
\be
\ket{C_n} = \frac{1}{2^{n/2}} \sum_{{\mathbf b}\in \{0,1 \}^n} (-1)^{\sum_i b_i b_{i+1}} \ket{{\mathbf b}}\,,
\ee
where the summation $\sum_i$ runs from $i=0$ to $i=n_b$ with $n_b=n-2$ for OBC and $n_b=n-1$, $b_n=b_0$ for PBC.  
%
%
It is an equal superposition of all $2^n$ computational basis states, where the phase of $|\mathbf{b}\rangle$
 is $-1$ if the number of neighboring $'11'$ pairs $(b_i = b_{i+1} = 1)$ is odd, and $+1$ otherwise. 
The phase is a direct consequence of the application of controlled-$Z$ (CZ) gates: 
 	\begin{eqnarray} \prod_{i=0}^{n_b}\mathrm{CZ}_{i,i+1}|\mathbf{b}\rangle=\prod_{i=0}^{n_b}(-1)^{b_ib_{i+1}}|\mathbf{b}\rangle,
 	\end{eqnarray}
Figure~\ref{fig:cluster} shows
the quantum circuit for a six-qubit cluster state with OBC. 

A minimal working example of  the end-to-end workflow for the classical shadow
protocol is shown in Fig.~\ref{fig:shadow-experiment}. After instantiating the
\texttt{ShadowUnveil} class, we invoke \texttt{.measure()} to perform the
randomized measurements. This method applies a  randomly sampled product of
single-qubit unitaries to the quantum state, followed by a projective
measurement in the computational basis. Upon completion, the random unitaries
and measurement outcomes are stored in the experiment registry
\texttt{experiment\_shadow.exps}.

\begin{figure*}[t]
        \begin{minipage}{\linewidth}
\begin{lstlisting}
import itertools
from qurry import ShadowUnveil
from qurry.recipe import cluster as Cluster
import numpy as np

def run_shadow_experiment():
    cluster_shadow = ShadowUnveil()
    n = 6
    cluster = Cluster(n, border_cond='open')
    cluster_exp = cluster_shadow.measure(cluster, 
                                snapshots=600, shots=400, tags=('cluster',))

    X = np.array([[0, 1], [1, 0]])
    Z = np.array([[1, 0], [0, -1]])
    subs01 = cluster_exp.analyze(selected_qubits=[0, 1], given_operators=[np.kron(Z, X)])
    print("The estimated entropy is ",subs01["purity","entropy"])
    print("The expectation value for X_0Z_1 is ",subs01["estimation","estimate_of_given_operators"])
\end{lstlisting}
        \end{minipage}
        \caption{Minimal usage of the \texttt{ShadowUnveil} class, using a cluster state as an example.}
        \label{fig:shadow-experiment}
\end{figure*}

\subsubsection{Pauli observables}
To estimate expectation values of operators, \texttt{.analyze()} accepts an optional \texttt{given\_operators} argument containing a list of operators represented as matrices. In this example, we estimate the expectation values of the stabilizers $X_0Z_1I_2$ and $Z_0X_1Z_2$ and one non-stabilizer $Z_0X_1I_2$ by supplying:
\begin{widetext}
	\begin{lstlisting}[numbers=none]
		ops = [np.kron(I, np.kron(Z,X)), np.kron(Z, np.kron(X, Z)),np.kron(I, np.kron(X,Z))]
		cluster_res = cluster_exp.analyze(selected_qubits=[0,1,2],given_operators=ops),
	\end{lstlisting}
\end{widetext}
where $X,Y,Z$ are Pauli operators, $I$ is the $2$ by $2$ identity matrix and the lower indices match that of the qubit indices in the quantum circuit. 
The tensor product for operators also follows the little-endian convention adopted by Qiskit, 
in which the rightmost factor acts on the lowest-index qubit. 
The output object of \texttt{.analyze()} is a custom class \texttt{analysis}. The expectation values can be extracted via:
\begin{lstlisting}[numbers=none]
cluster_res["estimation","trace_with_mean_rho_of_given_operators"]
\end{lstlisting}

The returned list is ordered to match the input \texttt{given\_operators}, so that the $i$-th entry corresponds to the expectation value of the $i$-th operator supplied. In this example, the outputs are 
\begin{widetext}
	\begin{lstlisting}
		# The output expectation values for XZI, ZXZ, ZXI are:
		[(0.9999999999999991+0j) (0.9999999999999998+0j) (0.0008148148148147683+0j)],
	\end{lstlisting}
\end{widetext}
agreeing with the analysis of the cluster states. 



\subsubsection{Overlap}
The next quantum property considered is the overlap of the cluster state with a set of
reference states. In this demonstation, we set periodic boundary for the
cluster state, achieved by appending a CZ gate connecting $q_0$ and $q_5$ in
the quantum circuit of Fig.~\ref{fig:cluster}.

Within the stabilizer formalism~\cite{Gottesman1997,Nielsen2010}, the density matrix of an $n$-qubit stabilizer state (such as the cluster state)
is given by~\cite{Aaronson2004,Audenaert2005}
\be
\rho^{[n]} = \prod_{i=0}^{n-1} (1+g_i) = \frac{1}{2^n} \sum_{G \in \mathcal{S}^{[n]}} G\,,
\label{eq:dm_stab}
\ee
where each element of the stabilizer group $\mathcal{S}^{[n]}$
is expressed as  $G=\prod_i g_i^{p_i}$ with $p_i \in \{0, 1\}$.
This formalism enables a convenient evaluation of the subsystem
overlap between the cluster state $\rho_C$ and another stabilizer state $\rho_\phi = \ket{\phi}\bra{\phi}$~\cite{Garcia2013}:
\be
\text{Tr}[\rho^{[n_A]}_C \rho^{[n_A]}_\phi] = \frac{1}{2^{n_A}}
\abs{\mathcal{S}^{[n_A]}_C \cap \mathcal{S}^{[n_A]}_\phi }\,,
\label{eq:overlap_stab}
\ee
where $n_A$ denotes the number of qubits in subsystem $A$,
and $\abs{\mathcal{S}_C \cap \mathcal{S}_\phi}$ represents the size of
the intersection of the corresponding stabilizer groups.

We consider the subsystem overlap between the cluster state and three other stabilizer states with PBC:
the all-zero state $\ket{0}^{\otimes n}$, the all-plus state $\ket{0}^{\otimes n}$,
and the Greenberger-Horne-Zeilinger (GHZ) state.
The total system size $n$ is taken to be even, and the subsystem $A$ is a contiguous interval
containing $n_A$ qubits, with $1 \le n_A \le n$.
Note that $|0\rangle^{\otimes n_A}$ and $|+\rangle^{\otimes n_A}$ are pure
reference states for any $n_A$, whereas the reduced state of a GHZ state on
any proper subsystem is mixed. Nevertheless, the overlap defined by
$\mathrm{Tr}(\rho\rho')$ applies to both cases.
The results are summarized in Table~\ref{tab:overlap}, with the derivations provided in Appendix~\ref{app:cluster}.


\begin{table}[t]
        \centering
        \begin{tabular}{|l|l|l|l|}
                \hline
                & $\ket{0}^{\otimes n_A}$ & $\ket{+}^{\otimes n_A}$ & $\ket{\text{GHZ}}$ \\ \hline
                $n_A = n$     & $1/2^{n}$   & $1/2^{n-2}$ & $1/2^{n-1}$ \\ \hline
                $n_A = n-1$   & $1/2^{n-1}$ & $1/2^{n-2}$ & $1/2^{n-1}$ \\ \hline
                $n_A = n-2$   & $1/2^{n-2}$ & $1/2^{n-2}$ & $1/2^{n-2}$ \\ \hline
                $n_A < n-2$   & $1/2^{n_A}$ & $1/2^{n_A}$ & $1/2^{n_A}$ \\ \hline
        \end{tabular}
        \caption{The overlap of an $n_A$-qubit subsystem with the cluster state under PBC.}
        \label{tab:overlap}
\end{table}

For a subsystem of $n_A \leq n$ qubits, the reduced density matrix is
calculated in \texttt{.analyze()} by assinging the qubit indices of the
subsystem in a list to the \texttt{selected\_qubits} argument. For example, for
the subsystem comprising $q_0$ and $q_1$, the argument is set to
\texttt{[0,1]}. The reduced density matrix estimated
as $\mathbb{E}[\hat{\rho}_A]$ corresponds to \texttt{mean\_of\_rho} returned by
\texttt{.analyze()}. The overlap between $\rho_A$ and an analytically known reference state $\sigma_A$ 
can then be estimated using Eq.~\eqref{eq:linoverlap}.

For the overlap estimater $\hat{f}_2$ in Eq.~\eqref{eq:f_overlap},
the projector $\Pi'$ is extracted by setting \texttt{use\_projector=True} in \texttt{.analyze()}.
The output field \texttt{average\_snapshots\_rho\_list}
then holds the projector in place
of the classical shadow snapshots normally returned by the function.

\begin{figure}[t]
 \centering
        \includegraphics[width=\linewidth]{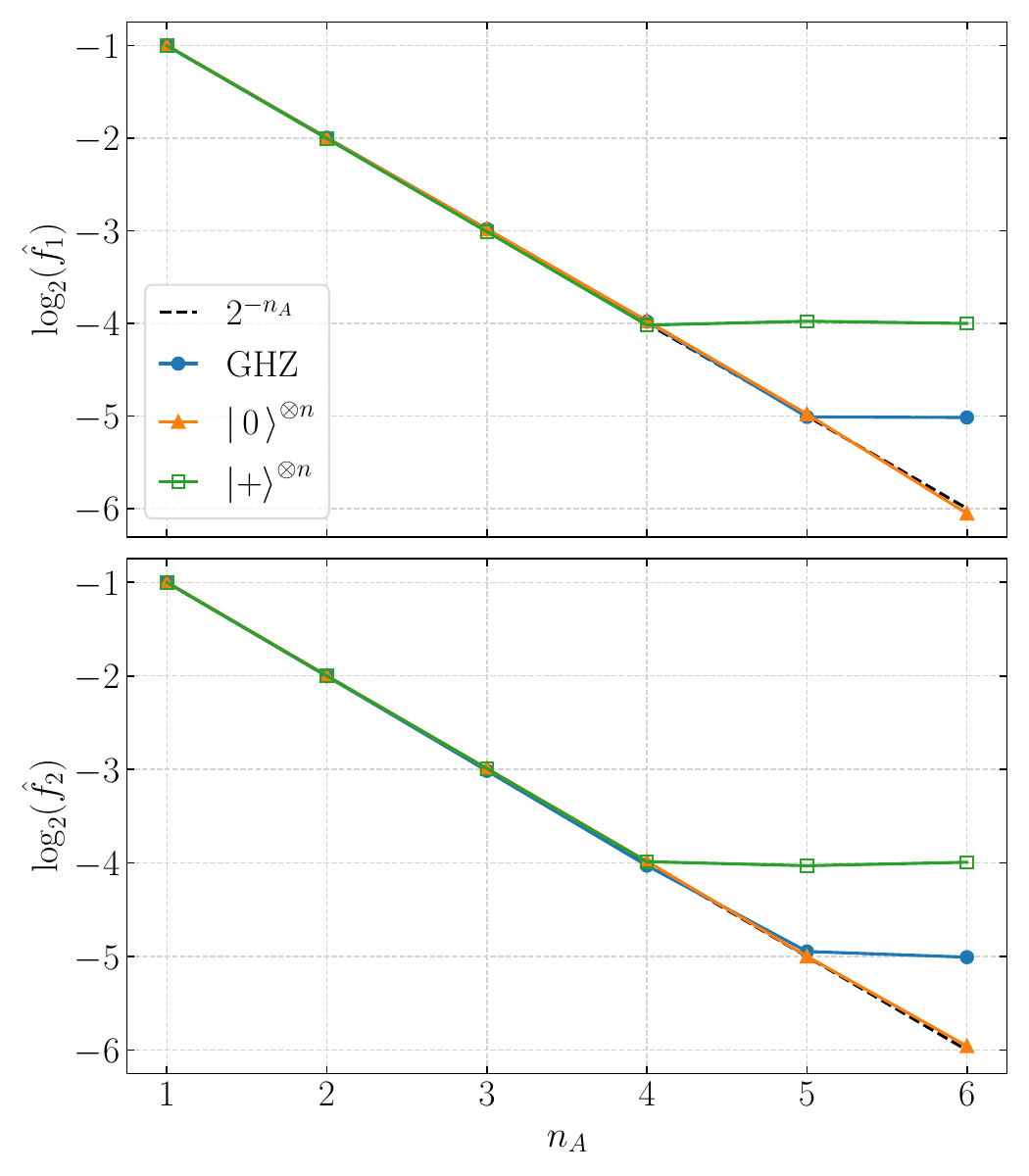}
 \caption{The overlap between the cluster state and GHZ, all-zero and all-plus states of subsystem size $n_A$,
estimated using Eq.~\eqref{eq:linoverlap} (top) and Eq.~\eqref{eq:f_overlap} (bottom).
The full system size is $n=6$ for all states.}
\label{fig:overlap}
\end{figure}

Figure~\ref{fig:overlap} shows the overlap of the cluster state with each
reference state as a function of the subsystem size $n_A$, with the total system size fixed at $n=6$.  
The results agree well with the theoretical values summaried in Table~\ref{tab:overlap}.

\subsubsection{Mutual information}

\begin{figure}[t]
        \centering
        \includegraphics[width=\linewidth]{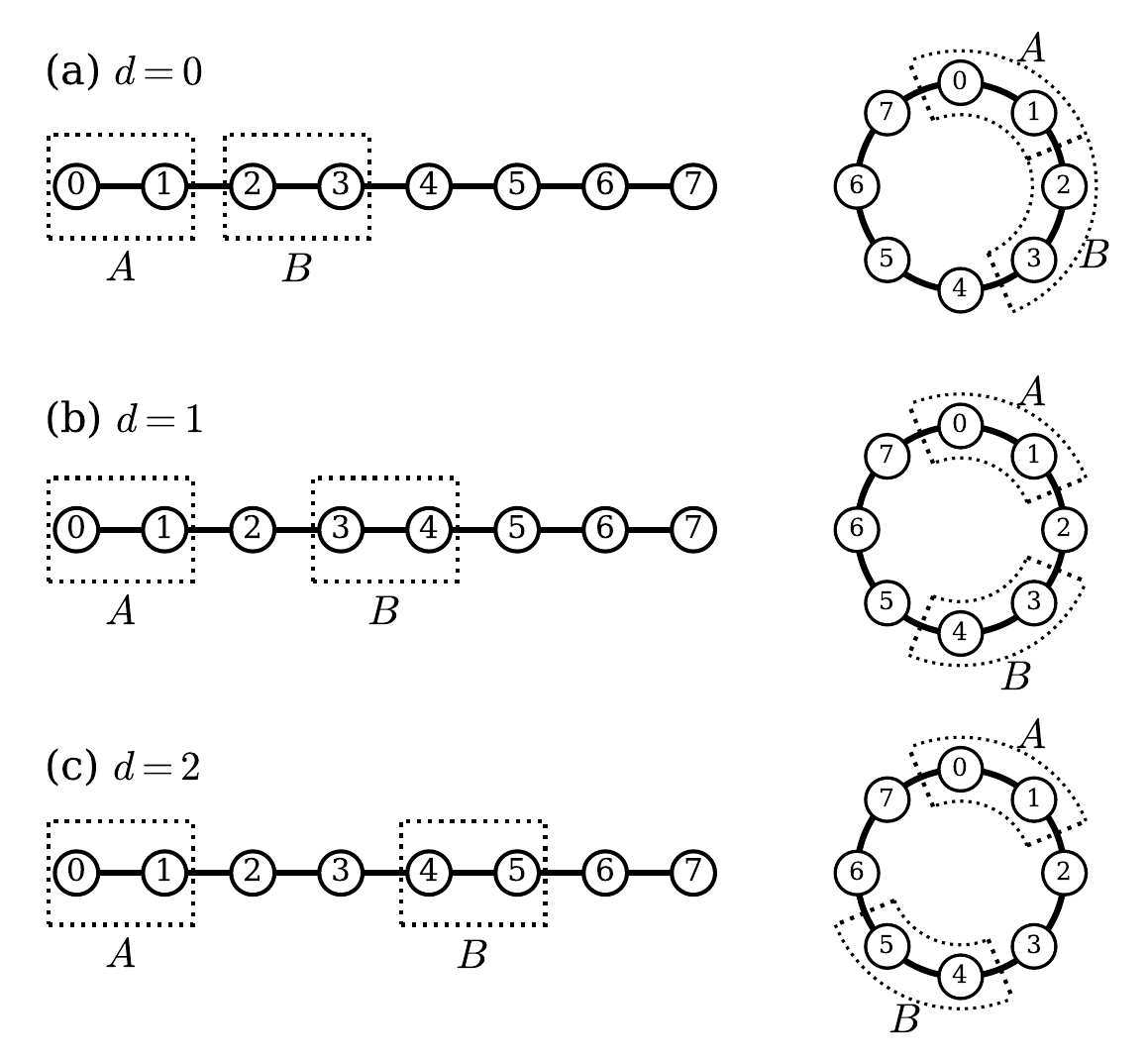}
        \caption{Illustration of the distance $d$ between two subsystems $A$ and $B$, enclosed by dotted lines, for an 8-qubit cluster state with OBC (left) and PBC (right). (a) $d=0$, (b) $d=1$, and (c) $d=2$. }
        \label{fig:midist}
\end{figure}
\begin{figure}[t]
        \centering
        \includegraphics[width=\linewidth]{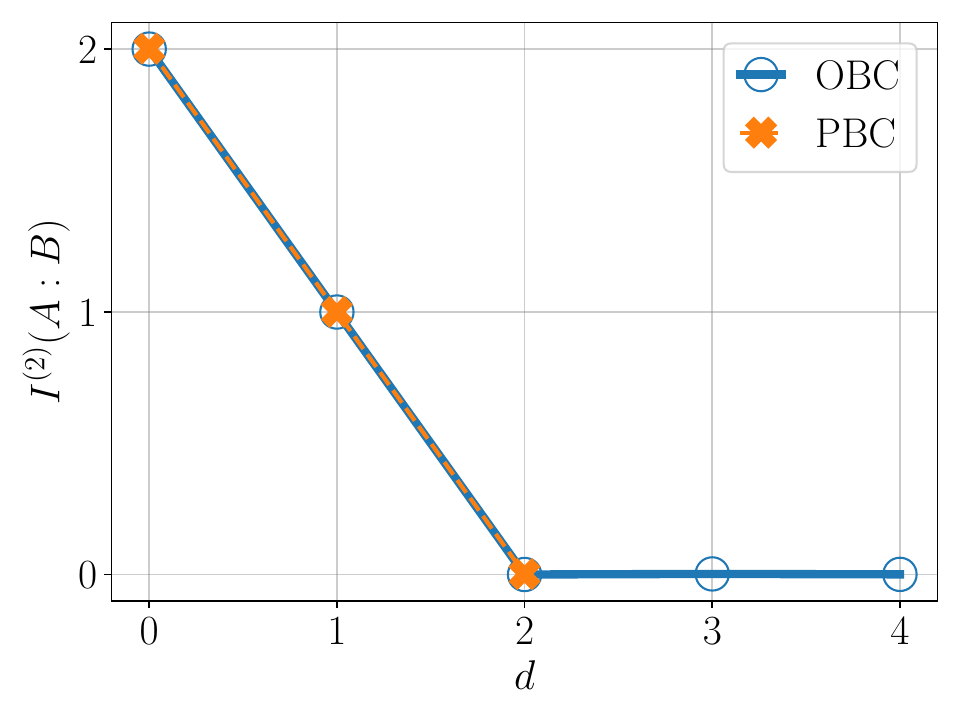}
        \caption{Second-order R\'enyi mutual information $I^{(2)}(A:B)$ as a function of the separation $d$ between subsystems $A$ and $B$ for the 8-qubit cluster state with OBC (blue circles) and PBC (orange crosses), simulated on \texttt{AerSimulator}. Parameters used: $N_U=3^8, N_K=100.$ }
        \label{fig:mi}
\end{figure}

We characterize the correlations between two contiguous subsystems $A$ and $B$ using the second-order
R\'{e}nyi mutual information~\cite{Choo2018}
\begin{equation}
	\label{eq:renyi_mi}
	I^{(2)}(A:B) = S^{(2)}(\rho_A) + S^{(2)}(\rho_B) - S^{(2)}(\rho_{AB}),
\end{equation}
where $S^{(2)}(\rho) = -\log_2 \mathrm{Tr}(\rho^2)$ is the second-order
R\'{e}nyi entropy, directly accessible from the purity $\mathrm{Tr}(\rho^2)$
estimated by the classical shadow protocol. 
For the stabilizer states considered here, the mutual information $I^{(2)}(A:B)$ is determined solely by the distance $d$
between the two subsystems, provided that both subsystem sizes satisfy $n_A, n_B>1$.
It can be evaluated via the following expression for the R\'enyi entropy~\cite{Hein2006}: 
\be
   S^{(2)}(\rho_x)=n_x - \abs{\mathcal{S}_x}\,,
   \label{eq:}
\ee 
where $x=A, B$ or $AB$, and $\mathcal{S}_x$ denotes the
the subgroup of stabilizer elements whose support is contained within $x$.  
The quantity $I^{(2)}(A:B)$ that characterizes the correlations between $A$ and $B$ vanishes 
if and only if $\rho_{AB} = \rho_A \otimes \rho_B$.

Within {\sf Qurrium},
the three purities, $\mathrm{Tr}(\rho_A^2)$, $\mathrm{Tr}(\rho_B^2)$, and
$\mathrm{Tr}(\rho_{AB}^2)$, required to evaluate Eq.~\eqref{eq:renyi_mi} are obtained by invoking \texttt{.analyze()} once per subsystem, for example:
\begin{lstlisting}[numbers=none]
	subA = cluster_exp.analyze(selected_qubits=[0, 1])
\end{lstlisting}
and analogously for $B$ and $AB$. The purity of each subsystem is then retrieved from the output of \texttt{.analysis()}, which returns the estimated $\mathrm{Tr}(\rho_x^2)$ with the key \texttt{('purity','purity')}. 

We evaluate the mutual information between two disjoint blocks, each consisting of two adjacent sites (i.e. $n_A=n_B=2$), 
as depicted in Fig.~\ref{fig:midist}. The distance $d$ between two
subsystems $A$ and $B$ is defined as the smallest number of lattice sites
separating the two subsystems. We consider three separations: $d=0$ where the
two subsystems share a boundary, $d=1$ where one qubit lies between the two
subsystems, and $d=2$ where two qubits separate them, as illustrated in panels
(a) to (c) in Fig.~\ref{fig:midist}, respectively. 

Figure \ref{fig:mi}  shows the simulation results for the 8-qubit cluster state on \texttt{AerSimulator}. For both boundary conditions, $I^{(2)}(A:B)$ decreases monotonically, taking values $2, 1, 0$ for $d = 0, 1, 2$ respectively, showing the short-ranged entanglement structure. For OBC, $I^{(2)}(A:B)$ remains $0$ at larger separations, since no stabilizer has support on both subsystems simultaneously for $d > 2$ \cite{Hein2006}. For PBC, the maximum distinct separation is $d = 2$ for the $8$-qubit ring with subsystem size $2$, as shown in the right panel of Fig.~\ref{fig:midist}(c). These results reflect the short-range nature of the three-body stabilizer interactions in the cluster state.

\subsection{Example:  Ising dynamics}
The Ising model is a paradigmatic model in condensed matter physics, describing
spin systems with pairwise ZZ interactions and capturing phenomena such as
quantum phase transitions and non-equilibrium dynamics \cite{Sachdev2011}. The
time evolution under this Hamiltonian is analytically tractable, making it an
ideal testbed for benchmarking quantum simulation on near-term
hardware~\cite{Kim2023}. In this example, we demonstrate {\sf Qurrium}'s
capability to track the time evolution of quantum properties across multiple
time steps, including the second-order R\'{e}nyi entropy, state overlap with
the initial state, and expectation values of selected operators. 

\begin{figure}[t]
		\centering
		\includegraphics[width=\linewidth]{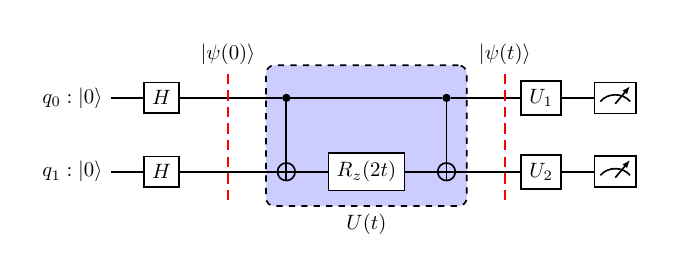}
		\caption{The quantum circuit for simulating a two-qubit Ising dynamics. The Hadamard gates are applied to prepare the initial state $|\psi(0)\rangle$. The colored region represents the unitary evolution operator. The quantum state is transformed to $|\psi(t)\rangle$ by $U(t)$. $U_1,U_2$ are random unitaries appended automatically in \texttt{.measure()} method.}
		\label{fig:isingcirc}
	\end{figure}

We consider the quench dynamics of a two-qubit system governed by the Ising Hamiltonian $H_\text{Ising}=JZ_0Z_1$,
starting from the initial state $|\psi(0)\rangle=|+\rangle|+\rangle$. We characterize the properties of the time-dependent state
$|\psi(t)\rangle=e^{-iJZ_0Z_1t}|\psi(0)\rangle$. The quantum circuit implementing this dynamics is shown schematically in 
Fig.~\ref{fig:isingcirc}.

In addition to noiseless simulation using Qiskit's
\texttt{AerSimulator} in the end-to-end workflow, we performed simulations
on the five-qubit superconducting quantum processing unit (QPU) fabricated by
Academia Sinica. 
Hereafter, we refer to this device as the AS QPU.
The AS QPU has a one-dimensional topology consisting of five physical qubits
connected by four tunable couplers. 
The experiment is implemented using Quantum Machines' OPX1000 controllers,
with execution scripts written in the pulse-level language QUA~\cite{QM2021}.
Further details of the experimental scripts are provided in Appendix \ref{sec:opx-classical-shadows}.

After collecting the measurement outcomes and corresponding random Pauli settings, 
we perform the data analysis using the \texttt{classical\_shadow\_complex} function
in the standalone estimation workflow in {\sf Qurrium}. 

\subsubsection{Pauli observables}

\begin{figure}[t]
        \includegraphics[width=\linewidth]{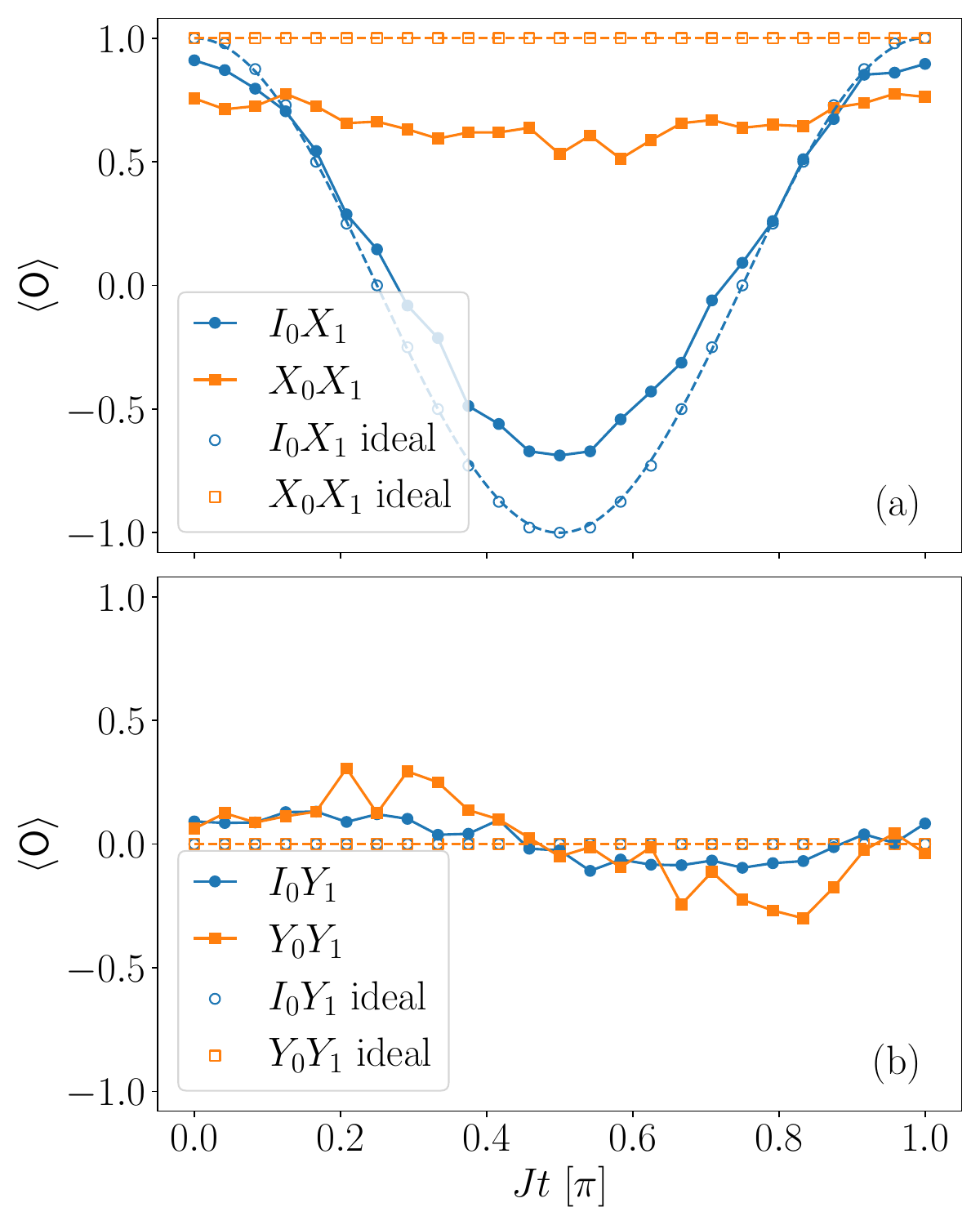}
	\caption{Estimated expectation values of the Pauli operators $I_0X_1$,
$X_0X_1$, $I_0Y_1$, and $Y_0Y_1$ with respect to the dynamical state under $ZZ$ Ising time evolution. 
The solid lines with filled markers are results from the AS QPU. 
The empty markers are results from Qiskit's \texttt{AerSimulator}. 
A total of $N_U = 90$ random Pauli settings are uniformly sampled from the set of tensor products of
single-qubit Pauli measurements. Each random Pauli setting is measured with
$N_K = 32$ shots. The dashed lines represent analytical results.}
        \label{fig:isingexp}
\end{figure}

We first evaluate the expectation values of Pauli operators as functions of time.
Since the Pauli-Z observable is conserved under the Ising Hamiltonian,
we focus on the dynamics of Pauli-X and Pauli-Y observables.
Unlike Larmor precession in a non-interacting system,
the $ZZ$ interaction in the Ising Hamiltonian constrains the dynamics of the Pauli observables.
While the $X$-component of an individual qubit
oscillates in time according to $\cos(2Jt)$, the corresponding $Y$-component
remains zero. Moreover, the two spins remain perfectly correlated throughout
the evolution, giving $\langle X_0X_1\rangle=1$ at all times.

Figure~\ref{fig:isingexp} presents the time-dependent expectation values of $I_0X_1$, $X_0X_1$,
$I_0Y_1$, and $Y_0Y_1$, obtained from both the AS QPU and noiseless simulation
using the AerSimulator (labeled ''ideal''). The QPU results exhibit deviations
from the ideal dynamics, while still capturing the qualitative behavior of the
observables.

\subsubsection{Overlap}

\begin{figure}[t]
        \includegraphics[width=\linewidth]{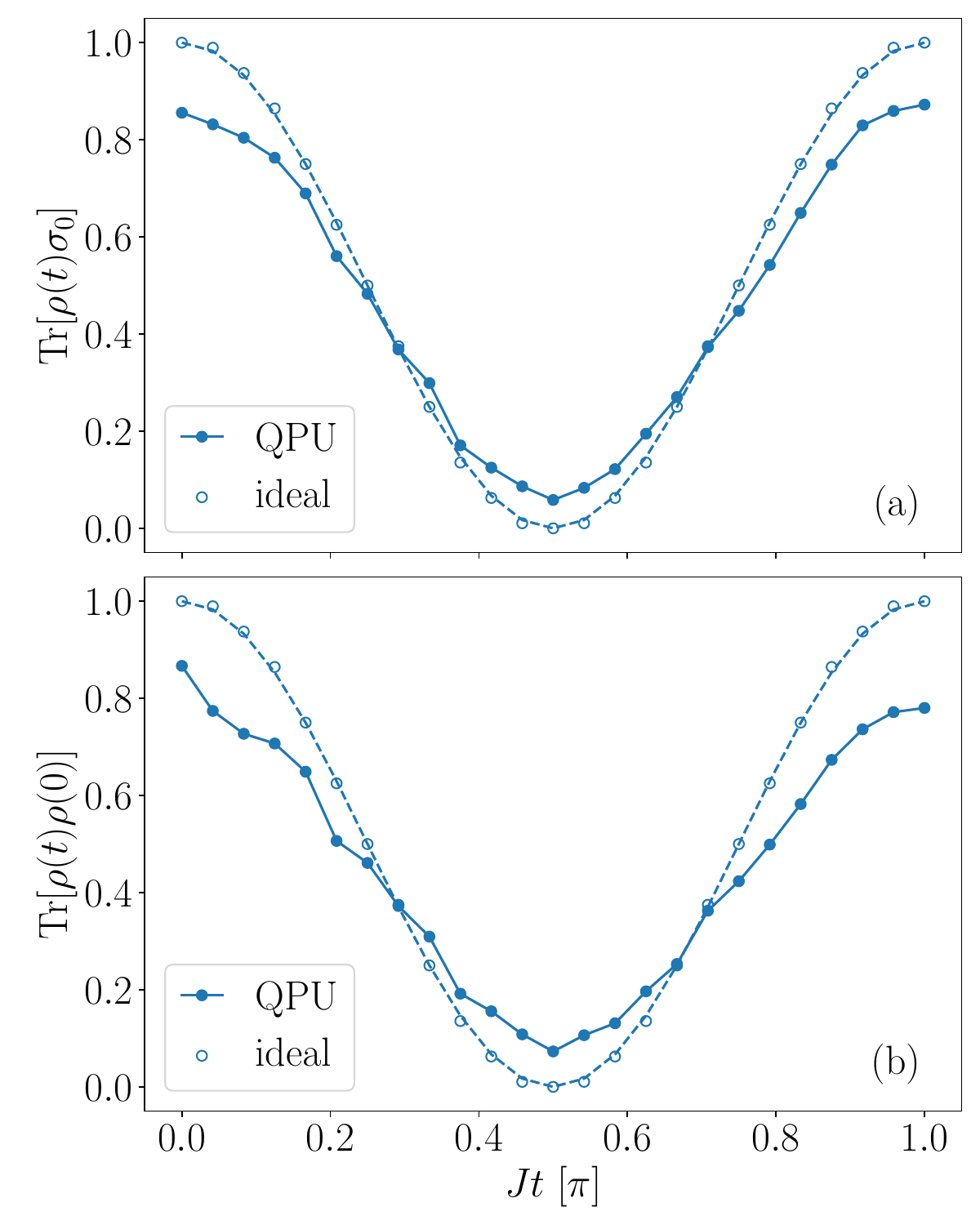}
	\caption{Estimated overlap between the dunamical state $\rho(t)$ under $ZZ$ Ising evolution
and the initial state $\rho(0)$, using $\hat{f}_1$ estimator (a), and
$\hat{f}_2$ estimator (b). The solid lines with filled markers are results from the AS
QPU. The empty markers are results from Qiskit's \texttt{AerSimulator}. 
A total of $N_U = 90$ random Pauli settings are uniformly sampled from the set of tensor products of
single-qubit Pauli measurements. Each random Pauli setting is measured with
$N_K = 32$ shots. The dashed lines represent analytical results.}
        \label{fig:isingoverlap}
\end{figure}

Next, we evaluate the overlap between the time-evolved state and the initial
state $|+\rangle^{\otimes 2}$, corresponding to the Loschmidt echo~\cite{HeylRev}.
Figure~\ref{fig:isingoverlap}(a) shows the overlap estimated using the linear estimator $\hat{f}_1(\rho,\sigma)$ in
Eq.~\eqref{eq:linoverlap}, where $\sigma$ is the exact density matrix of
the initial state. Figure~\ref{fig:isingoverlap}(b) shows the overlap estimated using the estimator
$\hat{f}_2(\rho, \rho')$ in Eq.~\eqref{eq:f_overlap}, where both $\rho$ and $\rho'$ are
reconstructed from measurement data. For the ideal simulations using
\texttt{AerSimulator} (empty markers), both estimators are in
good agreement with the analytical result $\cos^2(Jt)$ (dashed lines). 
For measurements performed on
the AS QPU, both estimators deviate from the analytical result, with
$\hat{f}_2$ showing a more significant deviation because the errors
associated with both density matrices accumulate in the overlap. 

At $Jt=\pi/2$, the vanishing overlap indicates that the time-evolved state
becomes orthogonal to the initial state. In the thermodynamic limit, such
zeros of the Loschmidt echo are associated with dynamical quantum phase
transitions~\cite{HeylPRL}.

\subsubsection{The second-order R\'{e}nyi entropy}

As a final example, we consider the purity and second-order R\'{e}nyi entropy of a single qubit and the full two-qubit system
as functions of time. 

\begin{figure}[t]
	 	\includegraphics[width=\linewidth]{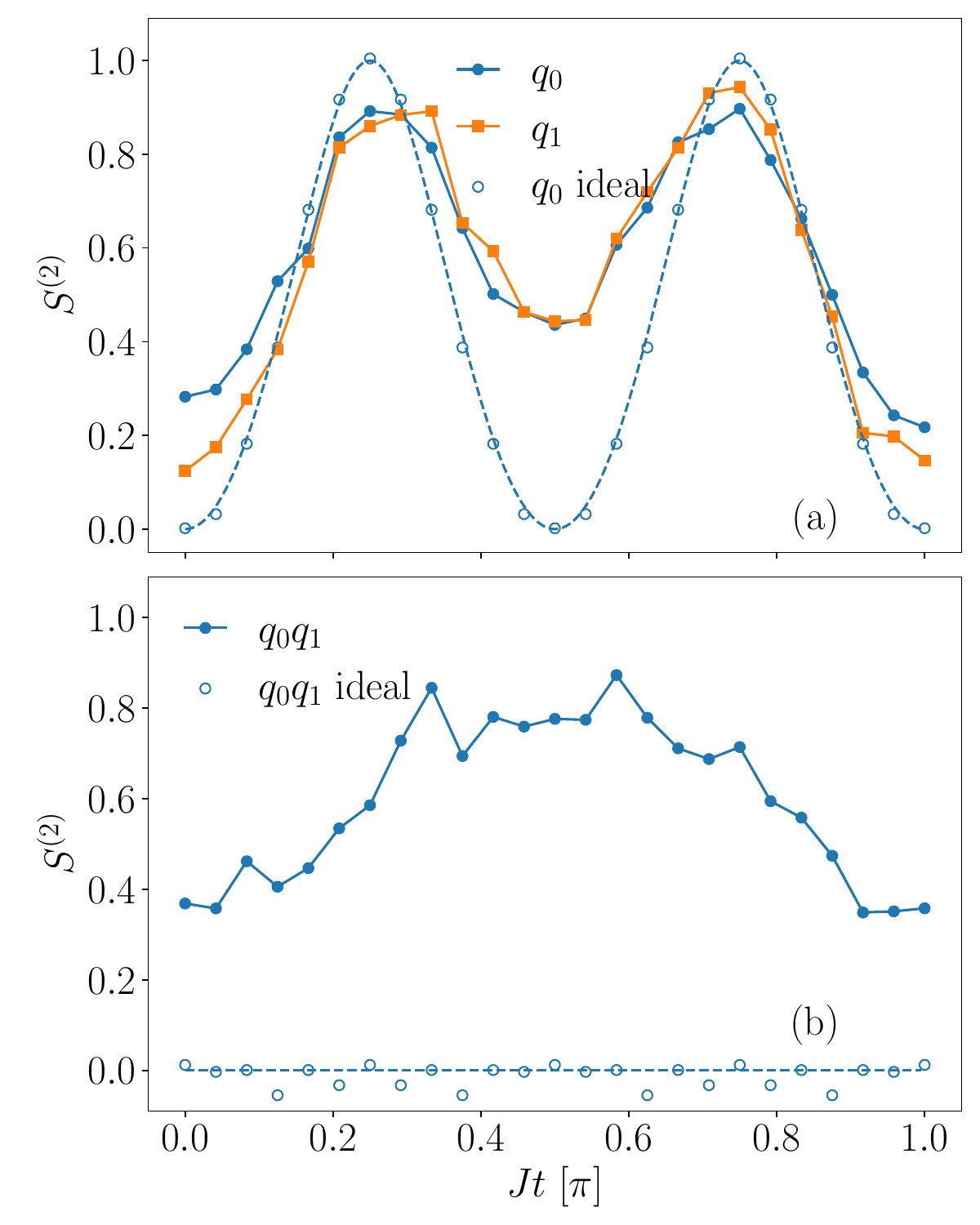}
		\caption{Estimated second-order R\'{e}nyi entropy of a single qubit (a) 
and the the full two-qubit system (b) under
$ZZ$ Ising evolution. The solid lines with markers are results from the AS QPU.
The empty markers are results from Qiskit's \texttt{AerSimulator}.
$N_U = 90$ random Pauli settings are uniformly sampled from the set of tensor products of
single-qubit Pauli measurements. Each random Pauli setting is measured with
$N_K = 32$ shots. The dashed lines represent analytical results.}
	 	\label{fig:isingentropy}
	 \end{figure}

The resulting second-order R\'{e}nyi entropy of the subsystems reveal the characteristic entanglement dynamics of the Ising model, as shown in Fig. \ref{fig:isingentropy} (a). The individual-qubit entropies oscillate away from unity, signalling the build-up of bipartite entanglement. The ideal simulation results (empty markers) agree well with the analytical solution $1-2\sin^2(Jt)\cos^2{(Jt)}$ (dashed lines). However, the subsystem entropy from AS QPU, shown by solid lines, does not reach the maximum of 1, suggesting
        the presence of amplitude damping or coherent gate errors
        that drive the state away from the maximally entangled
        configuration \cite{Nielsen2010}.
         Fig. \ref{fig:isingentropy} (b) shows the entanglement entropies of the entire system. The simulated results remain close to zero at all times steps as the full system evolves under a unitary. In contrast, the purity on AS QPU shows an overall larger value, showing the decoherence of the entire system.

\section{Conclusion and outlook}\label{sec:concl}
In this paper, we present {\sf Qurrium}, a Python package built on
Qiskit~\cite{qiskit2024}, that implements randomized measurement protocols for
estimating various quantum-state properties, including purity, second-order
R\'{e}nyi entropy, expectation values of observables, and state overlap, within the
classical shadow framework.  Furthermore, an unbiased overlap estimator
derived in this work extends the classical shadow protocol beyond the
estimation of linear functionals, enabling the comparison of quantum states
prepared independently on different hardware platforms from a single set of
randomized measurements.

We demonstrate two workflows provided by {\sf Qurrium}, namely the end-to-end workflow and the
standalone estimation workflow, through the estimation of various observables for
a cluster state and states undergoing Ising dynamics.  
The end-to-end workflow provides a convenient interface connecting
quantum circuit execution and data analysis, while the standalone estimation workflow
offers a versatile tool for analyzing data obtained from randomized measurements.
We further perform pulse-level simulations of quantum-quench dynamics on a superconducting-qubit processor developed by Academia Sinica and analyze the resulting measurement data using {\sf Qurrium}.

Several directions merit further development.
Possible extensions include:
\begin{description}[leftmargin=*]
\item[End-to-end hardware data acquisition.]	
Extending {\sf Qurrium} to support real hardware backends would enable users to execute randomized measurement circuits and collect measurement outcomes within the same software environment. This would consolidate the full workflow, from circuit construction and execution to data analysis. 
As demonstrated by our own OPX1000 implementation, the hardware execution can be addressed through a single program that pre-compiles the state-preparation circuit and generates the random Pauli settings in real time, or through the compilation of many distinct circuits (one per random Pauli setting) via the Qiskit runtime API~\cite{qiskit2024}. In the latter case, existing error mitigation methods~\cite{Cai2023}, including dynamical decoupling and Pauli twirling, are readily available through the Qiskit runtime API and could be directly applied to randomized measurement circuits.

\item[Integration of pulse calibration.]  The quantum state properties estimated by {\sf Qurrium}, including purity and state overlap, can serve as hardware-agnostic objective functions for pulse calibration and optimal control, naturally extending 
{\sf Qurrium}'s characterization capabilities toward direct hardware calibration. Incorporating the end-to-end workflow into a pulse optimization loop would enable calibration without requiring full quantum state tomography \cite{Krantz2019, Koch2022}, potentially reducing the calibration overhead on near-term quantum processors.


\item[Cross-platform verification protocols.] Developing standardized
cross-platform verification protocols would enable randomized measurement data
collected from different quantum-computing platforms to be analyzed and
compared within a common framework. Such protocols could facilitate the
benchmarking of measurement accuracy, hardware performance, and state-preparation quality across different devices \cite{Elben2020cross}, while taking advantage of the platform-independent nature of the standalone estimation workflow.
\end{description}

\section{Acknowledgement}
The authors acknowledge the Quantum Computing Test Space (QC-Test) at the Research Center for Critical Issues (RCCI), Academia Sinica, Taiwan for providing the quantum computing infrastructure and technical support. 
The authors acknowledge support from the National Science and Technology Council (NSTC) and the National Center for Theoretical Sciences (NCTS) in Taiwan. 
A.S. acknowledges useful discussions with Prof. Hui Khoon Ng on the implementation of the protocol. 
This work was supported under Grants No. 114-2119-M-007-013 and 115-2119-M-007-005.

\section{Author contributions}
H.-C.C is the main developer of the software. Y.-C.L. and H-.C.H. worked out theoretical and numerical results. T.-H.L. prepared well-calibrated superconducting QPU. A.S. designed the experimental control architecture and implemented the low-level OPX1000 programs used to execute the experiments on AS QPU. All authors contributed to the manuscript preparation and approved the final version of the manuscript.

Generative artificial intelligence was used to assist with grammar checking, and plotting scripts for workflow diagrams and figures. All scientific content and results were verified and approved by the authors, who take full responsibility for the accuracy of the manuscript.

\newpage
\onecolumn
\appendix
\section{Derivation of the overlap estimator}\label{app:overlap}
We estimate $\text{Tr}(\rho \rho')$ by pairing, for each  measurement setting $m$, the shot-averaged shadow of
$\rho$ with the shot-averaged raw projector of $\rho'$:
\be
\text{Tr}\bigl[\overline{\rho^{(m)}}\; \overline{\Pi'^{(m)}}  \bigr] =
\text{Tr}\left[\mathcal{M}^{-1}(\overline{\Pi^{(m)}} )\; \overline{\Pi'^{(m)}} \right]\,.
\ee
Denoting the measurement channel for a fixed setting $m$ by  $\mathcal{M}^{(m)}$,
we have
\be
\overline{\Pi^{(m)}} = \mathcal{M}^{(m)} (\rho)\,,
\ee
and
\be
\overline{\Pi'^{(m)}} = \mathcal{M}^{(m)} (\rho')\,.
\ee
Substituting these expressions back, we obtain 
\be
\text{Tr}\bigl[\overline{\rho^{(m)}}\; \overline{\Pi'^{(m)}}  \bigr] = 
\text{Tr}\bigl[ \mathcal{M}^{-1}(\mathcal{M}^{(m)} (\rho) ) \mathcal{M}^{(m)} (\rho')  \bigr]\,. 
\ee

Next, we average over the $N_U$ randomly chosen measurement settings.
It suffices to consider a single qubit, since the $n$-qubit measurement channel factorizes as the tensor product of the single-qubit channels. 
Using the standard Pauli expansion, we express the two single-qubit states as $\sigma=(I + \vec{r}\cdot \vec{P})/2$ and $\sigma' = (I + \vec{r}\,' \cdot \vec{P})/2$,
where $\vec{r}, \vec{r}\,'$ are the Bloch vectors and $\vec{P}=(X, Y, Z)$.
Applying the single-qubit measurement channel $\mathcal{M}_1^{(m)}$ for a fixed basis $P^{(m)}$ to the states $\sigma$ and $\sigma'$ yields
\be
\mathcal{M}_1^{(m)}(\sigma) = \frac{1}{2} \left( I + \text{Tr}(P^{(m)} \sigma) P^{(m)}  \right)\,,
\ee
and
\be
\mathcal{M}_1^{(m)}(\sigma') = \frac{1}{2} \left( I + \text{Tr}(P^{(m)} \sigma') P^{(m)}  \right)\,.
\ee
Applying the inverse channel $\mathcal{M}_1^{-1}(\,\cdot\,) = 3(\,\cdot\,) - I$ to $\mathcal{M}_1^{(m)}(\sigma)$, we obtain
\be
\mathcal{M}_1^{-1}\left( \mathcal{M}_1^{(m)}(\sigma)  \right) = \frac{3}{2}(I + \text{Tr}(P^{(m)} \sigma) P^{(m)}) -I =
\frac{1}{2} (I + 3\,\text{Tr}(P^{(m)} \sigma) P^{(m)}) \,.
\ee 
We then multiply the resulting correlated operators and take the trace, with $\text{Tr}(P^2)=\text{Tr}(I)=2$ and $\text{Tr}(P)=0$:
\begin{align}
	\text{Tr}\bigl[ \mathcal{M}_1^{-1}(\mathcal{M}_1^{(m)} (\sigma) ) \mathcal{M}_1^{(m)} (\sigma')  \bigr]
	& = \frac{1}{4}\text{Tr}\left[ \left( I + 3\,\text{Tr}(P^{(m)} \sigma) P^{(m)}  \right)  (I + \text{Tr}(P^{(m)} \sigma') P^{(m)})    \right] \nonumber \\
	& = \frac{1}{2} + \frac{3}{2} r_m r'_m\,,
	\label{eq:tr_right}
\end{align}
where $r_m \equiv \text{Tr}(P^{(m)} \sigma)$.
Averaging the expression in Eq.~\eqref{eq:tr_right} over the random Pauli basis, 
sampled with equal probability $1/3$ from $P^{(m)}\in \{X, Y, Z \}$, yields
\be
\left[ \text{Tr}\bigl[ \mathcal{M}_1^{-1}(\mathcal{M}_1 (\sigma) ) \mathcal{M}_1 (\sigma')  \bigr] \right]_\text{av}
= \frac{1}{2} + \frac{1}{2} \vec{r}\cdot \vec{r}\,'\,,
\ee
which is exactly $\text{Tr}(\sigma \sigma')$ for single-qubit states, since $\text{Tr}(\sigma \sigma')=\text{Tr}[(I+\vec{r}\cdot \vec{P})(I+\vec{r}\,'\cdot\vec{P})]/4 = (1+\vec{r} \cdot \vec{r}\,')/2$.

Upon taking the tensor product over all qubits for the $n$-qubit states $\rho$ and $\rho'$, we reproduce
\be
\left[ \text{Tr}\bigl[ \mathcal{M}^{-1}(\mathcal{M} (\rho) ) \mathcal{M} (\rho')  \bigr] \right]_\text{av} = 
\left[ \text{Tr}\bigl[\overline{\rho}\; \overline{\Pi'}  \bigr] \right]_\text{av} = \text{Tr}(\rho \rho')
\ee
and thereby verify that Eq.~\eqref{eq:f_overlap}
provides an unbiased estimator of $\mathrm{Tr}(\rho \rho')$: $\mathbb{E}\left[ \hat{f} [\rho,\,\rho'] \right ] = \text{Tr}(\rho \rho')$.

\section{Cluster state overlap in the stabilizer formalism}\label{app:cluster}

For an $n$-qubit system under periodic boundary conditions, a nontrivial cluster-state stabilizer 
$G_C = \prod_i g_i$ with $g_i=Z_{i-1}X_i Z_{i+1}$ can be rearranged as
\be
G_C = X_0 \left[ \bigotimes_{i=1}^{n-2} (-X_i) \right] X_{n-1} = (-1)^n \bigotimes_{i=0}^{n-1} X_i\,,
\label{eq:G_C}
\ee
by using the Pauli identities $Z^2_i=I$ and $Z_i X_i Z_i = -X_i$.
For even $n$, we can separate the chain into even and odd sublattices, yielding two stabilizers
\be
G_\text{even} = \prod_{i\; \text{even}} g_i = \bigotimes_{i\; \text{even}} X_i\,,
\label{eq:G_even}
\ee
and
\be
G_\text{odd} = \prod_{i\; \text{odd}} g_i = \bigotimes_{i\; \text{odd}} X_i\,,
\label{eq:G_odd}
\ee
whose product recovers $G_\text{even} G_\text{odd} = G_C$.

We now analyze the cluster-state stabilizers shared with the target states within a subsystem $A$.
\begin{itemize}
	\item The all-zero state $\ket{0}^{\otimes n}$ is stabilized by the generators $g_i=Z_i$ for all $i$.
	Therefore, the stabilizer intersection
	$\mathcal{S}_C^{[n_A]} \cap \mathcal{S}_0^{[n_A]}$ contains only the identity, yielding a subsystem overlap of $2^{-n_A}$.
	
	\item The all-plus state $\ket{+}^{\otimes n}$ is stabilized by the generators $g_i=X_i$ for all $i$.
	If the  subsystem $A$ covers the entire chain with even $n$, the stabilizer intersection contains four elements:
	\[ \mathcal{S}_C \cap \mathcal{S}_+ = \{I, G_\text{even}, G_\text{odd}, G_C \}\,, \]
	leading to
	$\text{Tr}[\rho^{[n]}_C \rho^{[n]}_+] = 4/2^n = 2^{-(n-2)}$. If $A$ contains $n_A = n-1$ qubits,
	the shared stabilizers with the cluster state are either $\{I,  G_\text{even} \}$ or $\{I, G_\text{odd}\}$,
	depending on whether the excluded qubit lies on an even or odd site. In both cases, this yields
	$\text{Tr}[\rho^{[n_A]}_C \rho^{[n_A]}_+] = 2/2^{n-1} = 2^{-(n-2)}$.
	For $n_A \le n-2$, any set of excluded sites necessarily includes both even and odd indices,
	so only the identity survives in the intersection, giving
	$\text{Tr}[\rho^{[n_A]}_C \rho^{[n_A]}_+]=1/2^{n_A}$.
	
	\item For $n$ qubits, the GHZ state $\ket{\text{GHZ}}= (\ket{0}^{\otimes n} + \ket{1}^{\otimes n} )/\sqrt{2}$
	is stabilized by the $n$ independent, commuting generators:
	\be
	g_0 = \bigotimes_{i=0}^{n-1} X_i,\quad g_j = Z_{j-1} Z_j\, \quad j=1, \cdots, n-1\,.
	\ee
	For $n_A=n$, the intersection $\mathcal{S}_C \cap \mathcal{S}_\text{GHZ} = \{I, g_0 \}$
	yields $\text{Tr}[\rho^{[n]}_C \rho^{[n]}_\text{GHZ}] = 2/2^n = 2^{-(n-1)}$.
	For $n_A \le n-1$, the intersection contains only the identity, giving a subsystem overlap of $2^{-n_A}$.
\end{itemize}

\section{Real-time classical shadows with the OPX1000}
\label{sec:opx-classical-shadows}
The OPX1000 control system is a modular, high-density quantum control platform designed by Quantum Machines. 
The Quantum Orchestration Platform (QOP)
is the control stack used to translate an experiment into precisely timed
operations on the quantum processor, while the OPX1000 is the real-time controller
that executes those operations.  Experiments are described in QUA, a language
that combines pulse-level commands with real-time variables, loops, arithmetic,
and control flow.  Consequently, the choice of measurement basis can be made
inside the running experiment, close to the quantum hardware, rather than by a
host computer that repeatedly constructs and submits a different circuit.

The guiding principle of our implementation is to keep the description of the
state separate from the machinery of the shadow protocol.  The user supplies
only a short gate-level state-preparation macro for \(\rho\).  Such a macro is a
sequence of logical instructions that is lowered to the calibrated pulse
operations of the selected qubits; in the present interface it can also be
generated from a Qiskit circuit.  The acquisition pipeline surrounds this macro
with the snapshot loop, basis randomization, measurement, state discrimination,
reset, and data streaming.  The same state-preparation description can therefore
be reused for different shadow sizes, numbers of repetitions, and local
measurement ensembles, without requiring the user to implement the randomized
measurement protocol itself.

Two controller capabilities make this construction both simple and scalable.
First, QUA supports real-time branching.  For every measured qubit \(i\), an
integer \(g_{m,i}\) labels the local rotation to be applied in the \(m\)th random
Pauli setting.  A controller-side \texttt{switch} statement selects the
corresponding gate from the configured single-qubit basis library before the
computational-basis readout.  For the Pauli ensemble this library implements the
rotations associated with the \(X\), \(Y\), and \(Z\) measurement axes.  The
optimized path uses conditional pulse plays for the two nontrivial rotations and
leaves the identity branch empty, but it is logically equivalent to the same
three-way switch.  Importantly, the branch is local: the controller evaluates
one choice for each qubit, rather than selecting one monolithic circuit from the
\(3^n\) possible tensor-product bases.

Second, the basis labels are generated at run time by the pseudo-random number
generator available to the OPX1000.  At the beginning of snapshot \(m\), the running
program samples
\begin{equation}
    g_{m,i} \in \{0,1,2\}, \qquad i=1,\ldots,n,
\end{equation}
independently for all qubits and stores the resulting vector
\(\mathbf{g}_m=(g_{m,1},\ldots,g_{m,n})\).  The state is then prepared, locally
rotated according to \(\mathbf{g}_m\), and measured \(N_K\) times.  Thus, in the
notation of Sec.~2, one random Pauli setting is sampled per snapshot and reused
for its \(N_K\) shots.  A seed makes the pseudo-random sequence reproducible,
while an optional table of preselected basis labels can be supplied when an
experiment requires a prescribed ensemble.  In the default mode, however, no
list of randomized circuits is generated or transferred by the host: every
local choice is produced and consumed by the controller during execution.

This organization turns the full acquisition into one persistent real-time
program,
\begin{equation}
\begin{gathered}
\underbrace{\text{prepare }\rho}_{\text{user macro}}
\quad\longrightarrow\quad
\underbrace{\text{sample }\mathbf{g}_m}_{\text{OPX1000}} \\
\underbrace{\text{rotate, measure, reset}}_{\text{OPX1000}}
\quad\longrightarrow\quad
\underbrace{(\mathbf{g}_m,\mathbf{b}_m)}_{\text{classical record}} .
\end{gathered}
\end{equation}
Here \(\mathbf{b}_m\) denotes the collection of bit strings obtained for the
\(N_K\) repetitions of setting \(m\).  Avoiding a host--controller round trip for
every setting reduces orchestration overhead and, more importantly, prevents
that overhead from growing with the number of tensor-product bases.  The
controller workload grows linearly with the number of qubits because the random
draw and branch are performed locally.  This is an implementation-level
efficiency: it does not alter the statistical sample complexity of the classical
shadow estimator or make full density-matrix reconstruction inexpensive.

The measurement record is streamed in a form that preserves the association
required by the inverse measurement channel.  For each repetition, thresholded
single-qubit outcomes are packed into an \(n\)-bit integer; the host-side result
layer converts these integers into bit-string counts for each snapshot.  In a
separate synchronized stream, the controller records the \(n\) basis labels
\(\mathbf{g}_m\).  
After collecting all measurement data, a lightweight adapter can map each label to its configured single-qubit unitary and pass the paired basis vectors and bit-string counts to
\texttt{classical\_shadow\_complex}.  Qurrium can then construct the snapshots
and estimate density matrices, purities, entropies, overlaps, or expectation
values of subsequently chosen observables.  The division of labor is therefore
clear: the OPX1000 performs timing-critical randomization and acquisition, whereas
Qurrium performs hardware-independent reconstruction and analysis.  Together
they realize the ``measure first, ask questions later'' philosophy without
requiring the user to manage the randomized measurement loop.

\newpage
\twocolumn
\bibliographystyle{quantum}
\bibliography{references}

\end{document}